\documentclass[prl,twocolumn,superscriptaddress,floatfix,amsmath]{revtex4-1}
\usepackage{graphicx}
\usepackage{color}
\usepackage{siunitx}
\usepackage{graphicx}
\usepackage[normal]{subfigure}
\usepackage[usenames,dvipsnames]{xcolor}
\usepackage[ansinew]{inputenc}

\usepackage{framed}
\usepackage{caption}
\usepackage{braket}

\newcommand{\be}{\begin{equation}}
\newcommand{\ee}{\end{equation}}
\newcommand{\dg}{^{\dag}}

\usepackage[colorlinks=true,urlcolor=blue,linkcolor=blue,citecolor=blue]{hyperref}

\begin{document}

\title{Chiral Quantum Optics}

\author{Peter Lodahl}
\author{Sahand Mahmoodian}
\author{S{\o}ren~Stobbe}
\affiliation{Niels Bohr Institute, University of Copenhagen, Blegdamsvej 17, DK-2100 Copenhagen, Denmark}
\author{Philipp Schneeweiss}
\author{J\"urgen Volz}
\author{Arno Rauschenbeutel}
\affiliation{Vienna Center for Quantum Science and Technology,
Atominstitut, TU Wien Stadionallee 2, 1020 Vienna, Austria}
\author{Hannes Pichler}
\affiliation{Institute for Quantum Optics and Quantum Information of the Austrian Academy of Sciences, 6020 Innsbruck, Austria}
\affiliation{Institute for Theoretical Physics, University of Innsbruck, 6020 Innsbruck, Austria}
\author{Peter Zoller}
\affiliation{Institute for Quantum Optics and Quantum Information of the Austrian Academy of Sciences, 6020 Innsbruck, Austria}
\affiliation{Institute for Theoretical Physics, University of Innsbruck, 6020 Innsbruck, Austria}

\date{\today}

\begin{abstract}
At the most fundamental level, the interaction between light and matter is manifested by the emission and absorption of single photons by single quantum emitters. Controlling light--matter interaction  is the basis for diverse applications ranging from light technology to quantum--information processing. Many of these applications are nowadays based on photonic nanostructures strongly benefitting from their scalability and integrability. The confinement of light in such nanostructures imposes an inherent link between the local polarization and propagation direction of light. This leads to {\em chiral light--matter interaction}, i.e., the emission and absorption of photons depend on the propagation direction and local polarization of light as well as the polarization of the emitter transition. The burgeoning research field of {\em chiral quantum optics} offers fundamentally new functionalities and applications both for single emitters and ensembles thereof. For instance, a chiral light--matter interface enables the realization of integrated non--reciprocal single--photon devices and deterministic spin--photon interfaces. Moreover, engineering directional photonic reservoirs opens new avenues for constructing complex quantum circuits and networks, which may be applied to simulate a new class of quantum many--body systems.
\end{abstract}

\maketitle

\section{Introduction}
The canonical setting of quantum optics is that of a single photon
interacting with a single quantum emitter  \cite{HarocheRaimondBook}. This elementary process underlies the essential physics of many phenomena and applications
of light\textendash matter interaction encompassing, e.g., photosynthesis,
vision, photovoltaics, optical communication, and digital imaging. A central goal of
quantum optics research is to develop tools for a complete control of light--matter interaction at the level of single quanta, while suppressing uncontrolled coupling to the surrounding environment, which would lead to decoherence \cite{QWI, *QWII}. Such a quantum optical toolbox
 provides the basis for applications in quantum communication and quantum--information processing. Here quantum--enhanced security and speed--up is realized by the aid of genuine quantum features such as quantum superposition or entanglement for encoding and processing information. Developing new elements of the toolbox  is a focus of ongoing research. The overarching goal of the research field is to assemble complex quantum circuits or quantum networks from basic quantum optical elements.

Chiral interfaces constitute an exciting new tool for the quantum mechanical control of the interaction between light and matter. Here, light--matter coupling inherently depends on the propagation direction of light and the polarization of the involved emitter transition. Consequently, a photon scattering off an emitter violates reciprocity, i.e., forward and backward propagation are different. In the most extreme case photon emission and absorption becomes {\em unidirectional}. Chiral coupling emerges naturally in nanophotonic structures such as waveguides and optical nanofibers, where light experiences tight transverse confinement. This introduces an inherent link between local polarization and propagation direction of light, which is a manifestation of spin-orbit coupling of light \cite{NanoOpticsBook,Bliokh2015NPHOT, Aiello2015NPHOT}. In conjunction with a polarization-dependent light--matter coupling strength, this leads to direction-dependent effects in the emission and absorption of photons, which can be described in the framework of {\em chiral quantum optics}. Beyond interest in the physics of chiral photon--emitter coupling {\em per se}, the latter has deep conceptual and practical implications.

\begin{figure}[ht]
	\includegraphics[width=\columnwidth]{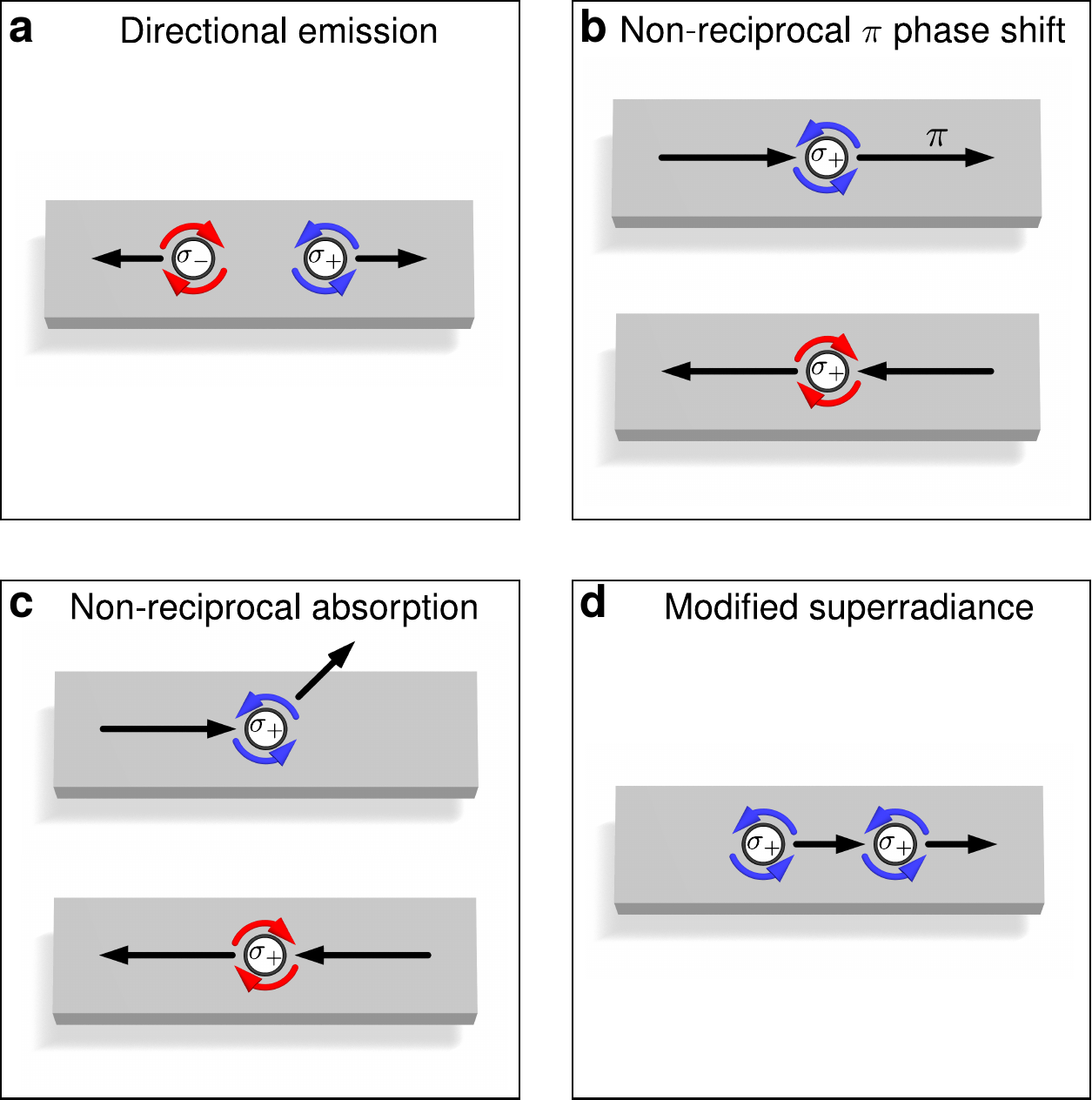}
	\caption{\label{Basic-concepts}
		\textbf{Illustration of basic chiral photon-emitter processes}.
		\textbf{a} Directional excitation of waveguide modes: Depending on the polarization of the dipole transition, $\sigma_+$ or $\sigma_-$, photons are emitted in one or the other direction into the waveguide.  \textbf{b} Strong chiral interaction between a photon and an emitter leading to a non-reciprocal phase shift of $pi$ for a forwardly propagation photon (upper figure) and vanishing interaction for a backwardly propagating photon (lower figure).
\textbf{c} Non-reciprocal absorption, i.e., direction-dependent scattering of photons into non-guided modes.
		\textbf{d} Chiral light--matter interaction in a multi-emitter case fundamentally modifies the collective emission.
		 Throughout the figure the colored arrows (blue/red) illustrate the local circular polarization of the electric field at the position of the emitter for the propagation direction indicated by the black arrows. }
\end{figure}

At the level of a single quantum emitter interacting with a pair of counter-propagating optical modes --- the paradigmatic case of a `one--dimensional (1D) quantum emitter' strived for in quantum optics \cite{Kimble1998PhysScripta} --- chiral coupling fundamentally modifies the elementary processes of light--matter interaction and, e.g., gives rise to directional photon emission, absorption, and scattering, see Fig.~\ref{Basic-concepts}. As a result, the direction-dependent light--matter interaction in chiral quantum optics enables non-reciprocal light propagation, i.e., light is prevented from re-tracing its forward path by the coupling to polarization-dependent scatterers, see Fig.~\ref{Basic-concepts}~(b,c). Based on this effect, nanophotonic optical isolators and circulators can be built that can be operated at ultra--low light levels down to single photons and be operated in a quantum superposition of simultaneously routing two different directions.

Even more intriguing are the implications of chiral quantum optics for quantum many-body systems that consist of 1D quantum emitters, which interact through the exchange of photons. Here directional photon emission implies a directional coupling between the quantum emitters via the waveguide: a first emitter sends out a photon, which is absorbed by a second emitter downstream, while photon emission from the second emitter can only be 'seen' by emitters located even further down the chain of emitters, see Fig.~\ref{Basic-concepts}~(d). On a formal level, chiral quantum optics allows to realize the paradigm of {\em cascaded quantum systems} \cite{Carmichael1993PRL,Gardiner:1993cy,QWI,QWII}, where a quantum system is directionally coupled to another quantum system without information backflow. Such unidirectional coupling has immediate applications in quantum information in order to achieve deterministic  quantum state transfer between qubits via a {\em chiral quantum channel}. From the perspective of solid--state or condensed--matter physics, the  non--reciprocal  photon--mediated interaction between emitters enables the implementation of a new class of unconventional quantum many--body systems. The dynamics of familiar reciprocal quantum many--body systems is often described solely by an interaction Hamiltonian, as in the case of dipole-dipole or Van der Waals interactions. In contrast, non--reciprocal interactions require a formulation in terms of a master equation, since these quantum systems are inherently open.

We  emphasize that chiral coupling of emitters to waveguides is distinct from the physics of  chiral edge channels, which were observed in recent seminal experiments in condensed--matter \cite{Hasan2010RMP} and photonic systems \cite{Hafezi2011NPHYS, Lu2014NPHOT, Raghu2008PRA, rechtsman2013strain, schine2016synthetic}. Chiral edge channels emerge as manifestations of topology in 2D materials leading to a topological protection of the wave transport through the channel towards backscattering from disorder.

\section{Physics of nanophotonic devices}

Nanophotonic devices control and confine the flow of light at a length scale smaller than the optical wavelength (400-700 nm for visible light). Recent years have witnessed remarkable progress in the ability to design and fabricate such nanophotonic structures and they are now widely employed in quantum optics as a powerful way of enhancing light--matter interaction \cite{Lodahl2015RMP}. In many nanophotonic devices, light is strongly confined transversely, i.e., in the plane orthogonal to the propagation direction. This generally leads to a component of the electric field, which oscillates along the direction of propagation \cite{Lax1975PRA}. According to Gauss's law, the longitudinal component is determined by the spatial variation of the transverse field components; see Box 1. The longitudinal and transverse field components oscillate $\pm \pi/2$~degrees out of phase with each other, where the sign depends on the propagation direction of the light; forward or backward. Thus, the electric field vector rotates with time meaning that the light field is elliptically polarized and therefore carries spin angular momentum. In contrast to paraxial waves, the spin in general has a component transverse to the propagation direction \cite{Bliokh2012PRA}. The direction dependence of the relative phase between the longitudinal and transverse fields implies that the transverse spin component flips sign with the inversion of propagation direction \cite{Bliokh2015NPHOT, Aiello2015NPHOT}. This locking of  transverse spin and propagation direction of confined light leads to chiral effects in the interaction of light with polarized emitters. The properties of light in transversely confined geometries and the concept of transverse spin are introduced in more detail in Box 1.

\begin{figure*}[t]
\includegraphics[width=0.8\textwidth]{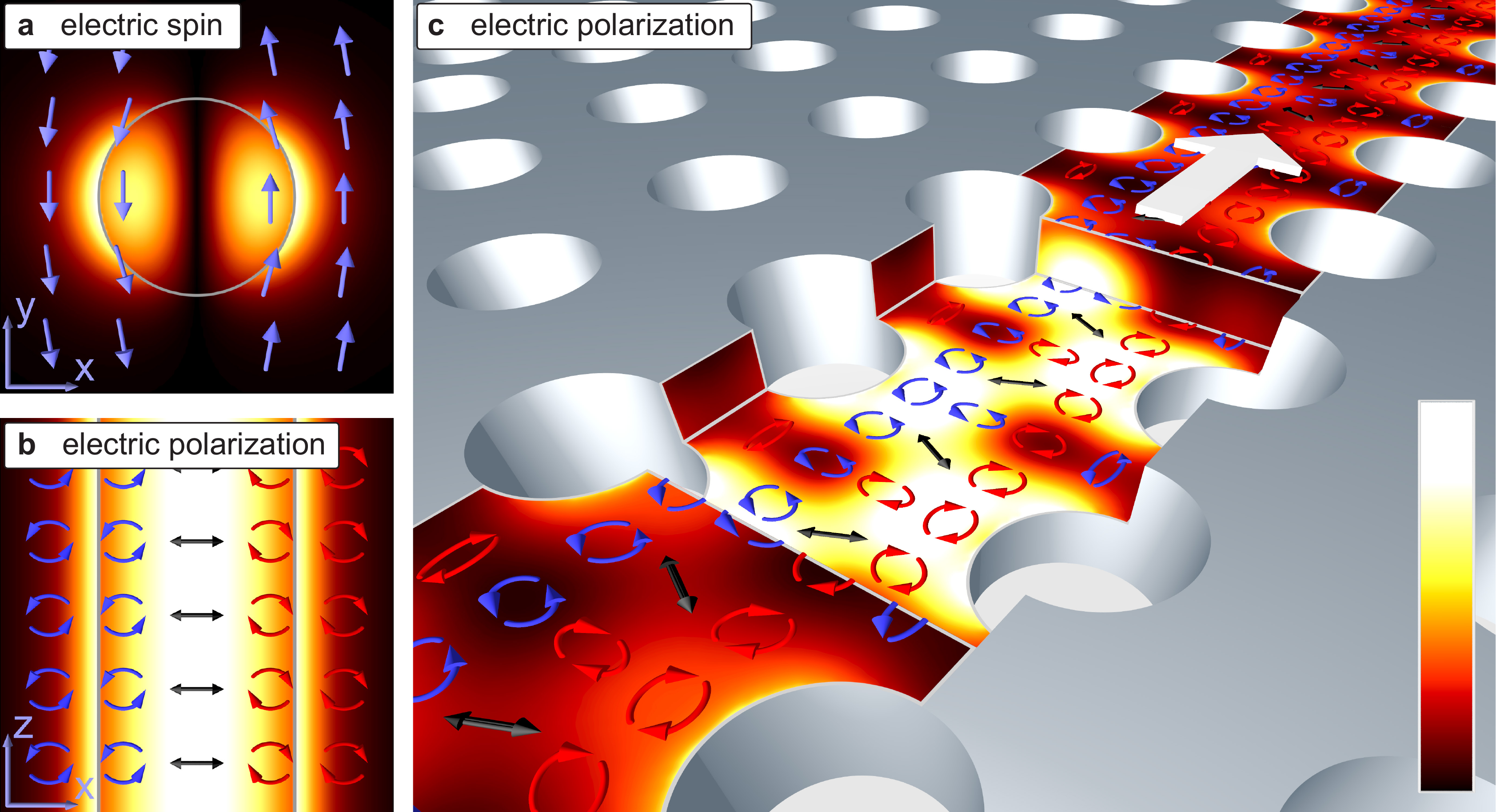}
\caption{\label{fig:Fiber+GPW} \textbf{Electric field polarization and spin in optical nanofibers and waveguides}.   \textbf{a} Color map of the electric spin density $(\left| \boldsymbol{\mathcal{S}}_{\mathcal{E}} \right|)$ in the cross-sectional plane of the fiber with arrows indicating its direction. \textbf{b} Color map of the field intensity with the local polarization represented by blue, red, and black arrows to indicate the the two circular in-plane polarizations and linear polarization, respectively. The considered radius of the fiber is \SI{250}{\nano\metre}, wavelength \SI{852}{\nano\metre}, refractive index 1.45, and the propagation direction is along +z.  \textbf{c} Color map of the electric field intensity and local polarization of a photonic-crystal glide-plane waveguide \cite{Sollner2015NNANO}. The grey arrow shows the propagation direction and the inset shows the linear color scale from zero (dark colours) to maximum (light colours) scaled field intensity. The applied lattice constant is \SI{240}{\nano\metre}, membrane thickness \SI{160}{\nano\metre}, the wavelength  is  \SI{911}{\nano\meter}, and the membrane refractive index 3.46.}
\end{figure*}

In order to illustrate the polarization properties of transversely confined light fields, Fig.~\ref{fig:Fiber+GPW} shows the local electric field polarization for a subwavelength-diameter optical fiber and a glide-plane photonic-crystal waveguide. In the optical fiber, light confinement relies on total internal reflection while in the photonic waveguide it is induced by Bragg scattering due to periodic modulations of the refractive index. When a linearly polarized laser beam is launched into an optical nanofiber, the polarization of the nanofiber-guided light remains linear on the fiber axis. However, the polarization is nearly circular close to the nanofiber surface and in the evanescent field where the strong field gradients give rise to a strong longitudinal component of the electric field \cite{LeKien2004OPTCOMM}; see Fig.~\ref{fig:Fiber+GPW} \textbf{b}.
In the photonic-crystal waveguide the local polarization can be engineered in order to maximize the transverse spin component at spatial positions where the field intensity of the guided mode is high. The glide-plane waveguide is developed for this purpose and Fig.~\ref{fig:Fiber+GPW} \textbf{c} illustrates the local polarization and field intensity in this case. The transverse spin is exemplarily shown in Fig.~\ref{fig:Fiber+GPW} \textbf{a} for a nanofiber, which depicts the electric part of the spin density as a function of the transverse position.

\begin{figure*}[ht]
	\includegraphics[width=\textwidth]{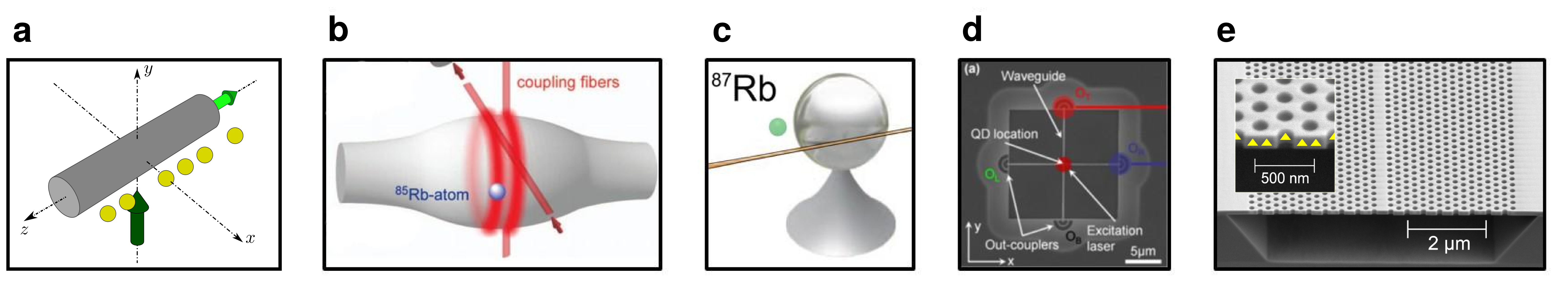}
	\caption{\label{Fig:systems} \textbf{Nanophotonic devices used for chiral coupling between light and quantum emitters}. \textbf{a} Optical nanofiber with an array of cold atoms trapped in the evanescent field surrounding the fiber. Figure reproduced from \cite{Mitsch2014NCOM}.  \textbf{b} and \textbf{c} Whispering-gallery-mode microresonators that confine light by total internal reflection along the circumference and provide strong coupling to a single rubidium atom. Figures reproduced from \cite{Junge2013PRL} and \cite{Shomroni2014Science}. \textbf{d} Cross of two nanowire planar waveguides with guided modes coupled to a quantum dot placed in the crossing region. Figure reproduced from \cite{Luxmoore2013PRL}. \textbf{e} 2D photonic-crystal membrane with an introduced waveguide obtained by leaving out a row of holes of the photonic lattice. A single layer of quantum dots is embedded in the center of the membrane. Figure reproduced from \cite{Sapienza2010Science}.}
\end{figure*}

The emergence of transverse spin of light and accordingly spin-momentum locking is universal for evanescent and other strongly confined optical fields \cite{Bliokh2015Science, VanMechelen2016Optica}. The associated directional coupling has been observed experimentally in both dielectric \cite{Luxmoore2013PRL, Junge2013PRL, Luxmoore2013APL, Petersen2014Science, Neugebauer2014NL, Rodriguez-Fortuno2014ACS, Shomroni2014Science, Mitsch2014NCOM, Sollner2015NNANO, leFeber2015NCOMM,  Coles2016NCOM} and plasmonic nanostructures \cite{Lee2012PRL, Lin2013Science, Rodriguez2013Science, OConnor2014NCOM} by coupling both classical \cite{Petersen2014Science, Neugebauer2014NL, Rodriguez-Fortuno2014ACS, leFeber2015NCOMM, Lee2012PRL, Lin2013Science, Rodriguez2013Science, OConnor2014NCOM} and quantum \cite{Luxmoore2013PRL, Junge2013PRL, Luxmoore2013APL, Shomroni2014Science, Mitsch2014NCOM, Sollner2015NNANO, Coles2016NCOM} emitters to the confined light fields. The robustness of the chiral points against unavoidable fabrication imperfections in photonic-crystal waveguides has been assessed \cite{Lang2015PRA, Coles2016NCOM} and chiral coupling is a well characterized and robust phenomenon, which is readily implementable in a range of applications.
Figure~\ref{Fig:systems} presents a selection of nanophotonic devices featuring chiral coupling, including atoms in the vicinity of tapered optical fibers and microresonators (Fig.~\ref{Fig:systems} \textbf{a}-\textbf{c}) and quantum dots in photonic waveguides (Fig.~\ref{Fig:systems} \textbf{d}-\textbf{e}).

\begin{framed}
\section{Box 1 | Confined light and transverse photon spin}

When confining light transversely to the propagation direction, longitudinal field components occur naturally. Consider exemplarily a focused light beam of wavelength $\lambda$ and angular frequency $\omega$ that propagates along $\pm z$. The electric field is ${\bf E}(\boldsymbol r, t)=\boldsymbol{ \mathcal{E}}(\boldsymbol r) /2e^{-i(\omega t\mp k z)}+c.c.$, where $\boldsymbol{ \mathcal{E}}$ is the complex amplitude and $k=2\pi/\lambda$ the wavenumber. Gauss' law for a slowly varying amplitude along $z$ leads to
\begin{equation}
\mathcal{E}_z=\mp\frac{i}{ k}\left(\frac{\partial \mathcal{E}_x}{\partial x}+\frac{\partial\mathcal{E}_y}{\partial y}\right),
\label{eq:Elong}
\end{equation}
i.e., the longitudinal component becomes comparable to the transverse components once the latter vary significantly on the length scale of $1/k=\lambda/2\pi.$ The $\mp\pi/2$ phase of $\mathcal{E}_z$ implies that even a linearly polarized beam becomes elliptically polarized in the strongly focused region and that the sense of rotation of ${\bf E}$ depends on the direction of propagation $\pm z$.
The same line of reasoning also applies to the magnetic field $\boldsymbol{\mathcal H}$.

The spin angular momentum density of the light field, $\boldsymbol{\mathcal{S}}= \boldsymbol{\mathcal{S}}_{\mathcal{E}} + \boldsymbol{\mathcal{S}}_{\mathcal{H}}$, has an electric part, $\boldsymbol{\mathcal{S}}_{\mathcal{E}}=-\frac{i\epsilon_0}{2\omega}\boldsymbol{\mathcal{E}}^* \times \boldsymbol{\mathcal E}$, and a magnetic part, $\boldsymbol{\mathcal{S}}_{\mathcal{H}}=-\frac{i\mu_0}{2\omega}\boldsymbol{\mathcal{H}}^* \times \boldsymbol{\mathcal H}$, with the vacuum permittivity $\epsilon_0$ and vacuum permeability $\mu_0$, respectively. Throughout this manuscript, we limit our discussion to electric dipole transitions and therefore only consider the properties of $\boldsymbol{\mathcal{S}}_{\mathcal{E}}$.
The presence of $\mathcal{E}_z$ leads to a transverse spin component, $\boldsymbol{\mathcal{S}}_{\mathcal{E}}^{\rm trans}$, and the sign change in Eq.~(\ref{eq:Elong}) implies that $\boldsymbol{\mathcal{S}}_{\mathcal{E}}^{\rm trans}$ flips sign when the propagation direction is reversed.
This inherent link between  the transverse spin and the propagation direction, or spin-momentum locking, is a consequence of time-reversal symmetry of Maxwell's equations. We note that a similar behavior can be obtained in a cavity supporting two degenerate transverse modes of orthogonal circular polarization, i.e., spin--momentum locking can be emulated \cite{Coles2014OE}. In this case, the cavity must support two degenerate circularly-polarized modes; in the waveguide geometry the degeneracy of the forward and backward propagation modes is enforced by time--reversal symmetry.

Fig. \ref{Fig:prism} illustrates the emergence of transverse spin for the simple case of total internal reflection at a dielectric interface.
The ratio of the longitudinal and transverse field components is given by $\mathcal{E}_z/\mathcal{E}_y=-i [ 1-(n_2/n_1)^2/\cos^2\theta]^{1/2}$ and leads to nearly circular polarization for large angles of incidence $\theta$. In this case, the tip of the electric field vector describes an almost circular trajectory (circular arrow) in the $x$-$z$-plane. The electric spin density is thus fully transverse, $\boldsymbol{\mathcal{S}}_{\mathcal{E}}=\boldsymbol{\mathcal{S}}_{\mathcal{E}}^{\rm trans}$, and points along the  $+y$ direction. For grazing incidence on a silica-vacuum interface, the expectation value of the photon spin is $\langle \hat{ \boldsymbol S} \rangle= 0.96\,\hbar\,\boldsymbol e_y$, i.e., the photon spin is close to the maximum  value given by Planck's constant $\hbar$.

\includegraphics[width=0.9\columnwidth]{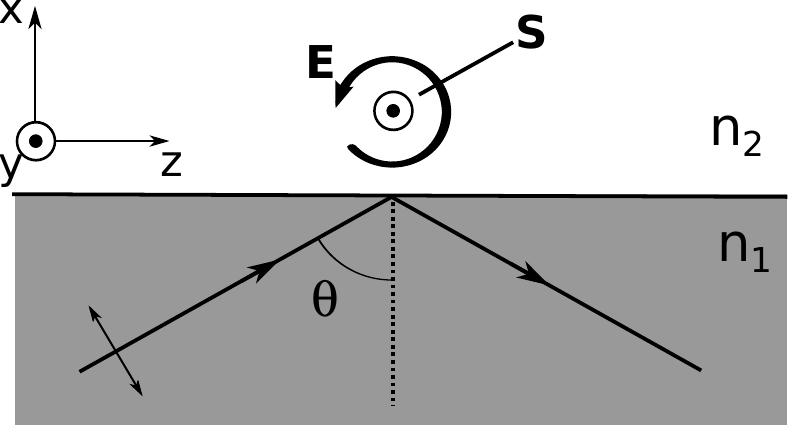}
\captionof{figure}{Total internal reflection at a dielectric interface between two media with refractive indices $n_1>n_2$. An incident $p$-polarized light field undergoing total internal reflection leads to an evanescent field in medium 2. }
\label{Fig:prism}

\end{framed}


\section{Chiral light-matter interaction}
\label{chap:LightMatter}

We will now turn to the effect of transverse spin and spin-momentum locking on the interaction of a single quantum emitter and a pair of counter-propagating optical modes -- the 1D emitter. Given that the elementary processes of light-matter interaction, emission, absorption, and scattering, depend strongly on the angular momentum of the light, fundamental changes in the physics of light-matter interaction are anticipated.  The coupling is described in the following within the dipole approximation, which is sufficient in most experimental situations. However, it should be mentioned that the extended size and lack of parity symmetry encountered in the case of quantum dots mean that they may be sensitive also to the magnetic fields \cite{Tighineanu2014PRL}. We emphasize that the present analysis could readily be extended to include multipolar interaction terms as well.

The light--matter interaction  strength can be quantified by the $\beta$-factor, i.e., the ratio between the rate of spontaneous emission into the $\pm$ modes and the total emission rate  including all other modes. The forward and backward emission rates are $ \gamma_{\pm} \propto \left| {\boldsymbol{d}^*}\cdot \boldsymbol{\mathcal{E}}_{\pm} \right|^2$, where $\boldsymbol{d}$ and $\boldsymbol{\mathcal{E}}_\pm$ are the complex emitter dipole matrix element and the amplitude of the forward and backward mode, respectively. Time--reversal symmetry of Maxwell's equations demands $\boldsymbol{\mathcal{E}}_+ = \boldsymbol{\mathcal{E}}_-^*$. In the presence of transverse spin, this immediately implies that the polarizations of the two counter-propagating modes differ, meaning that $\boldsymbol{\mathcal{E}}_+ \neq \boldsymbol{\mathcal{E}}_-$. As a consequence, the forward and backward emission rates are in general not symmetric, $ \gamma_{+}\neq\gamma_{-}$, and the $\beta$-factor is direction-dependent; see Box 2 for a detailed introduction of the $\beta$-factor.

\begin{figure*}
	\includegraphics[width=0.8\textwidth]{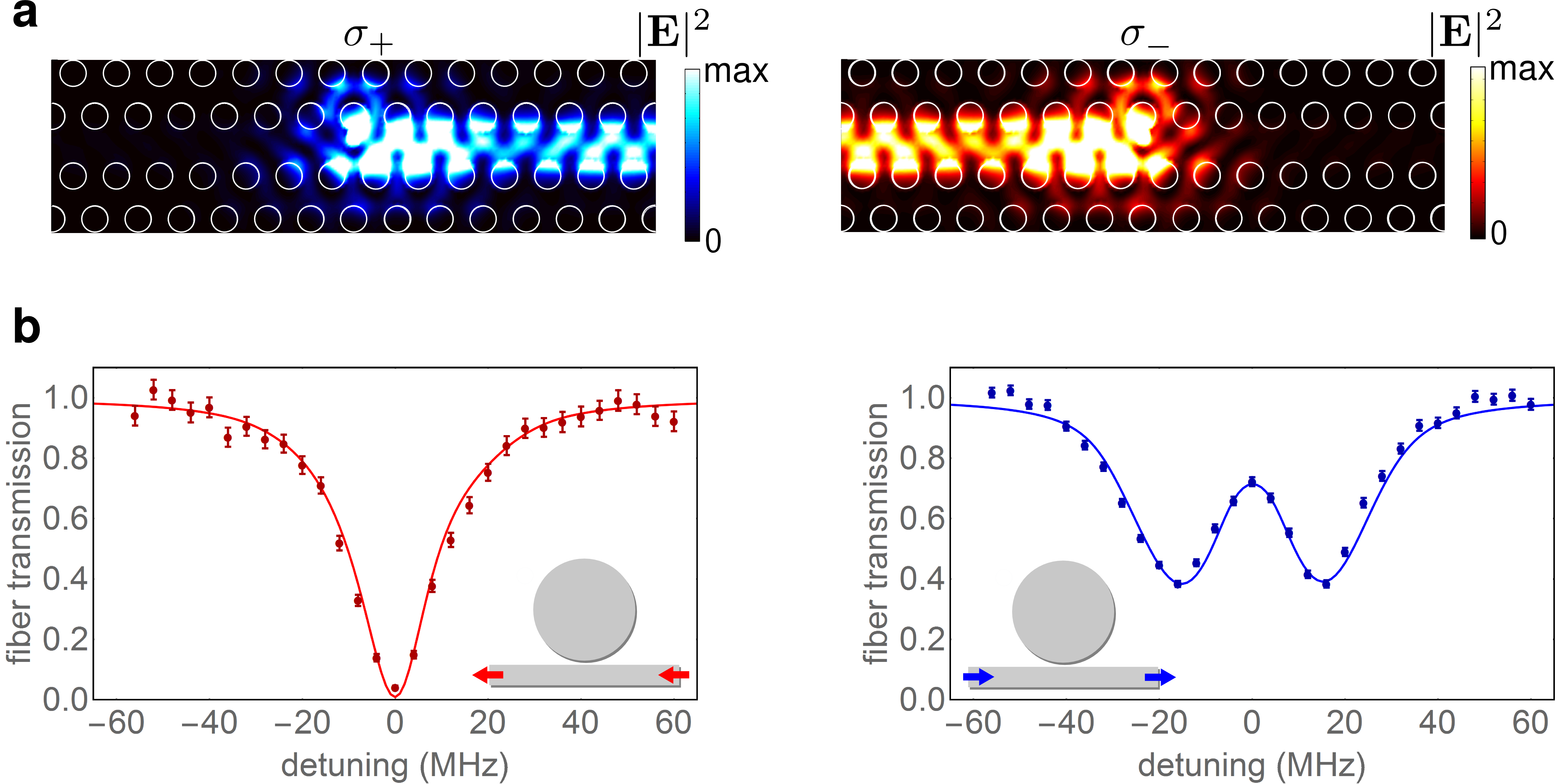}
	\caption{\textbf{Examples of chiral light-matter interaction in photonic nanostructures.}
		\textbf{a} Directional spontaneous decay of two opposite circularly ($\sigma_{\pm}$) polarized dipoles into opposite directions of a glide-plane photonic-crystal waveguide.  The plot illustrates the spatial distribution of the emitted intensity. Figure reproduced from Ref.~\cite{Sollner2015NNANO} . \textbf{b} Direction-dependent coupling between a single atom and nanofiber guided light. The atom couples strongly to the light propagating in forward direction (blue transmission spectrum) and only weakly to backward propagating light (red transmission spectrum).
		Data from Ref.~\cite{Junge2013PRL}.
	}
	\label{fig:DirChan}
\end{figure*}

In the ideal case the polarizations at the position of the emitter, $\mathcal{E}_+$ and $\mathcal{E}_-$ are circular and thus orthogonal for opposite propagation directions.  Therefore, a circularly polarized dipole emitter matched to the polarization of one propagation direction solely emits along this direction. This is the basic physical principle of chiral coupling leading to directional single-photon emission: a circularly polarized dipole emits preferentially along one direction in the waveguide depending on the helicity of the transition. This phenomenon is demonstrated in Fig.~\ref{fig:DirChan} \textbf{a} for a photonic--crystal waveguide. Directional emission has been observed experimentally with a range of different types of emitters embedded in various photonic nanostructures~\cite{Luxmoore2013PRL, Junge2013PRL, Luxmoore2013APL, Petersen2014Science, Neugebauer2014NL, Rodriguez-Fortuno2014ACS, Shomroni2014Science, Mitsch2014NCOM, Sollner2015NNANO, leFeber2015NCOMM,  Coles2016NCOM, Lee2012PRL, Lin2013Science, Rodriguez2013Science, OConnor2014NCOM}.

Two different coupling regimes are identified depending on the coupling strength between the emitter and the pair of counter-propagating modes. If these modes are defined by free-space beams or homogeneous dielectric waveguides, chiral effects occur when the emitter is positioned away from the beam center. Consequently, coupling to non-guided modes is substantial leading to $\beta\ll1$, i.e., the emitter dynamics is largely determined by the non-guided modes. Nonetheless the residual coupling to the waveguide modes may still be fully chiral, e.g., $\gamma_+\neq 0$ and $\gamma_-= 0$. For $\beta\gtrsim1/2$ the emitter dynamics is dominated by the interaction with the waveguide modes, and in the extreme case of $\beta \sim 1$ a fully deterministic single-photon--single-emitter interface is obtained. This regime may be reached by suppressing the coupling to non-guided modes by exploiting photonic band-gap effects \cite{Arcari2014PRL} and highly-efficient directional coupling has been reported ~\cite{Sollner2015NNANO, leFeber2015NCOMM}. An alternative approach is to enhance the coupling of the single emitter to the waveguide via an optical resonator exhibiting chiral coupling~\cite{Junge2013PRL, Shomroni2014Science}.

\begin{framed}
	\section{Box 2 | Chiral photon emission and scattering}

The interaction  strength between a single emitter and a pair of counter-propagating modes is quantified by the directional $\beta$-factor. It is given by the ratio of the spontaneous emission rate into the
$\pm z$-directions, $\gamma_\pm$, and the total emission rate,
	\begin{equation}
	\beta_\pm=\frac{\gamma_\pm}{\gamma_++\gamma_-+\Gamma}~,
	\end{equation}
	where $\Gamma$ is the emission rate into all other modes. The total coupling efficiency is $\beta=\beta_++\beta_-$. In the presence of transverse spin, the emission into the counter-propagating modes is in general not symmetric, i.e., $\beta_+ \neq \beta_-$ corresponding to directional coupling.
	
The scattering of guided light on quantum emitters in waveguides is also strongly modified by chiral coupling. The scattering amplitude transmission $t$, and reflection $r$ coefficients for a narrow--band optical pulse on resonance with the emitter, are in the weakly saturated regime given by
	\begin{eqnarray}
	t_\pm & = & 1-2\beta_\pm, \label{eq:t} \\
	r_\pm & = & -2\sqrt{\beta_+\beta_-}~. \label{eq:r}
	\end{eqnarray}
As a special case these equations describe also the scattering of a narrow--band single--photon pulse. Striking qualitative differences between the symmetric and chiral emitter--waveguide coupling are found.
First, consider deterministic coupling $(\beta=1)$ and symmetric coupling $(\beta_+=\beta_-=1/2)$. Here, the emitter acts as a perfect mirror for the guided light, i.e., $|r_\pm|^2=1$. In contrast, in the ideal chiral case of, e.g., $\beta_+=1$, $\beta_-=0$, the emitter is perfectly transparent, i.e., $|t_\pm|^2=1$.  Furthermore, the interaction imprints a direction-dependent phase shift of 0 or $\pi$ onto the light propagating along $-z$ and $+z$, respectively.
	
The absorption of the guided light also differs strikingly for symmetric and chiral coupling. It is given by the fraction of emission being scattered by the emitter into non-guided modes: $A_\pm=  1-|t_\pm|^2-|r_\pm|^2.$ For symmetric coupling, at most 50 \% of the guided light can be absorbed, i.e., $A_\pm \leq 0.5$, and maximum absorption is reached for $\beta=0.5$. In contrast, for ideal chiral coupling and $\beta = 0.5$, one obtains perfect absorption of, e.g., $A_+=1$, while light that is propagating in the opposite direction is fully transmitted, $A_-=0$.
	
\end{framed}

%

\begin{figure}[h!]
\includegraphics[width=0.9\columnwidth]{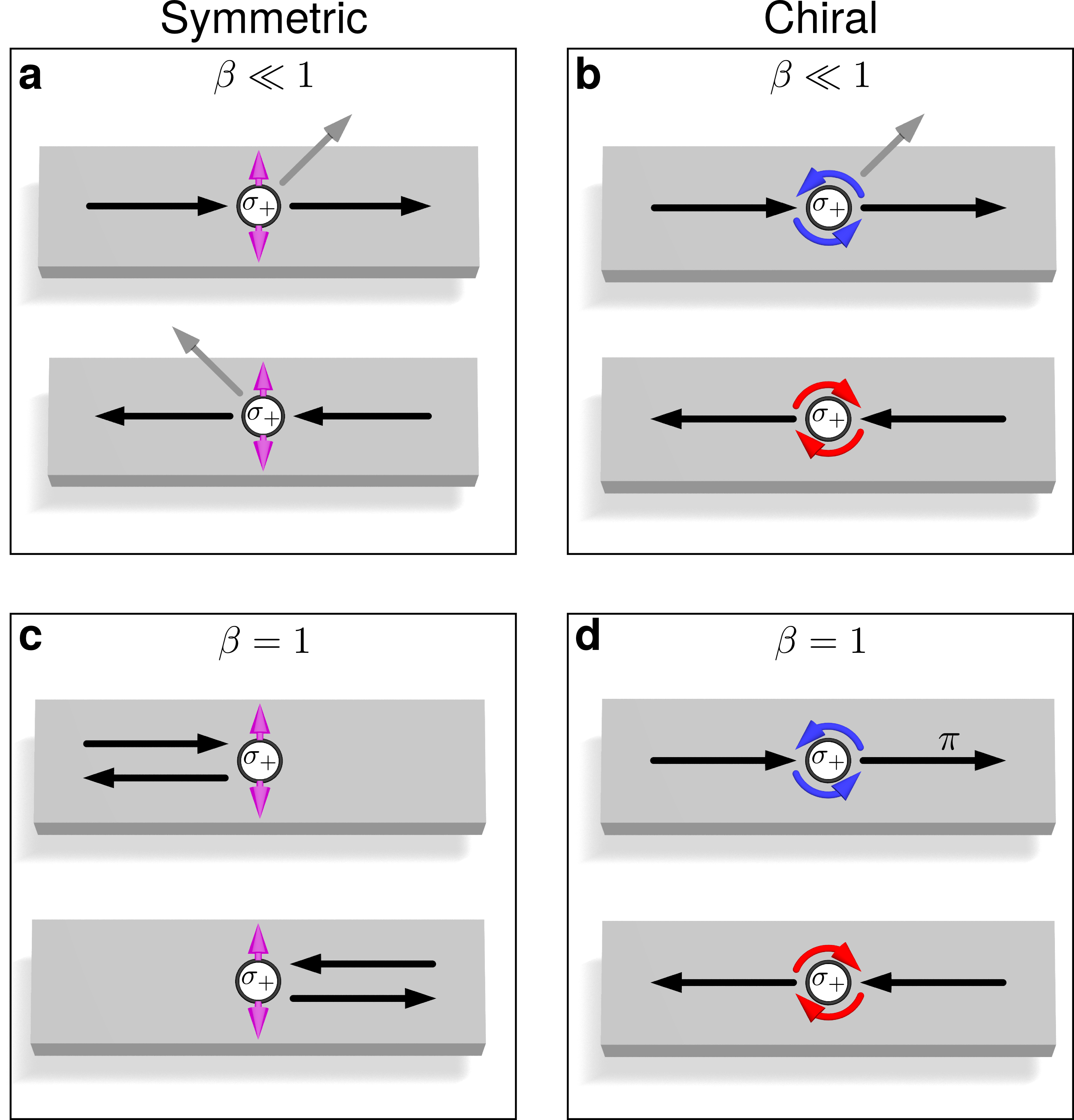}
\captionof{figure}{\label{fig:regularVchiral}\textbf{Photon-emitter scattering for symmetric and chiral coupling.} Illustration of the case of $\beta \ll 1$(a and b)  and deterministic coupling $\beta = 1$ (c and d) for the symmetrically-coupled case (left column) and chiral coupling (right column) comparing both forward (upper figures) and backward (lower figures) propagation.  The different effects are elaborated in the main text. The pictorial representations are identical to the ones introduced in Fig. \ref{Basic-concepts}. }
\end{figure}

 Chiral quantum optics results in strongly direction-dependent reflection and absorption of the guided light, cf. Box 2 for further details. Figure~\ref{fig:DirChan}~\textbf{b} shows an example, where the presence of a single atom results in a strong directional dependence of the waveguide transmission~\cite{Junge2013PRL}. Figure~\ref{fig:regularVchiral} illustrates photon--emitter scattering for symmetric and chiral coupling both in the case of small and large coupling efficiency. In the  symmetric cases, reciprocal photon transmission and reflection is obtained, while chiral coupling breaks reciprocity of the photon transport in the waveguide.
 The weak coupling regime can be used, e.g., to engineer the amount of light being absorbed from the waveguide by using the coupled emitter to scatter a fraction of the guided light.
In the the most extreme case of deterministic coupling (cf. Fig.~\ref{fig:regularVchiral} \textbf{c} and \textbf{d}) where chiral coupling turns an otherwise perfectly reflecting emitter into a fully transparent emitter, which imparts a nonreciprocal $\pi$ phase shift on the light.

Chiral light--matter coupling also occurs in the interaction with detuned light fields. In this case, a propagation direction-dependent phase shift arises when the light interacts with an emitter. In turn, the interaction with the optical fields leads to an AC Stark shift of the emitter's energy levels~\cite{Grimm2000AdvAtMolOpt}. For elliptically polarized fields this leads to a Zeeman-state-dependent shift ~\cite{Cohen-Tannoudji1972PRA}. Consequently, the transverse spin results in a position- and propagation--direction dependent artificial magnetic field ~\cite{Lembessis2014JOSAB,  Mochol2015SciRep}, which has been proposed as a way of realizing an optical analogue of a wire trap for cold atoms~\cite{Schneeweiss2014NJP}. Moreover, such fields have been employed experimentally for the spatially selective quantum-state preparation of cold atoms~\cite{Mitsch2014PRA}.

\begin{figure}[t]
\includegraphics[width=\columnwidth]{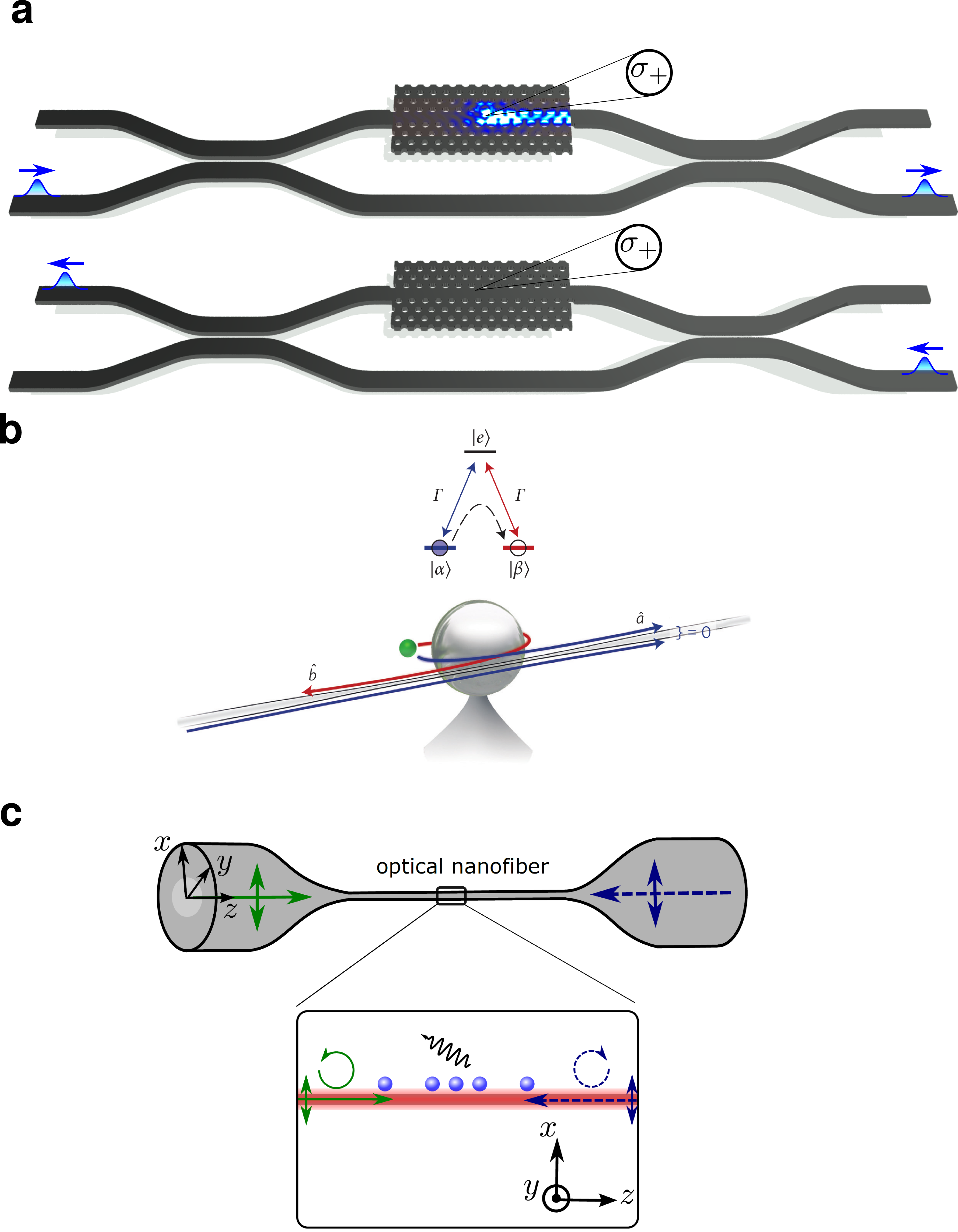}
\captionof{figure}{\label{Quantum-applications} \textbf{Applications of chiral light-matter interaction.} \textbf{a} Layout of an optical circulator for single photons based on non-reciprocal phase shift in a Mach-Zehnder interferometer containing a glide-plane waveguide with a quantum emitter. Figure reproduced from~\cite{Sollner2015NNANO}.
 \textbf{b} The chiral interaction in a $\Lambda$-type system can be employed to realize a single-photon controlled optical switch. Figure reproduced from~\cite{Rosenblum2016NPHOT}. \textbf{c} Illustration of atoms, which scatter guided light from a nanofiber with direction-dependent rates leading to the realization of an optical diode.  }
\end{figure}

\section{From Basic Functionalities to Quantum Many-Body Systems}

Chiral interfaces enable a range of new applications, such as controlling the flow of light with nanophotonic devices, and can be assembled to realize complex integrated optical circuits. In particular the intrinsic quantum character of chiral light-matter interaction and the ability to control it at the level of single quanta leads to unique properties and functionalities. Immediate applications in, e.g.,~quantum information processing and for  constructing complex quantum networks are anticipated. Moreover, chiral interfaces can be used to engineer unique non-equilibrium quantum many-body system of photons and 1D emitters. In the following, we first discuss the recent experimental and conceptual achievements concerning basic elementary devices based on chiral light-matter interaction. We then outline the prospects and applications of chiral quantum optics in the context of quantum many-body physics.

\subsection{Elementary Devices Based on Chiral-Light Matter Interaction}



The possibility to break Lorentz reciprocity is a fundamental ingredient for designing devices such as optical isolators and circulators ~\cite{Jalas2013NPHOT}. In conventional non-reciprocal optical devices, this is induced by a magnetic field in conjunction with  magneto-optical materials, the time modulation of the optical properties of the system, or an optical nonlinearity. Non-reciprocity that relies on chiral coupling to break Lorentz reciprocity, however, can be based solely on the atomic spin, which is in general associated with a polarization-dependent coupling strength to realize optical isolators and circulators~\cite{Lenferink2014OE,Xia2014PRA}.
 Optical isolators and circulators utilizing chiral coupling to achieve non-reciprocal absorption or phase shifts have been demonstrated with either a small ensemble of atoms coupled to an optical nanofiber, cf. Fig.~\ref{Quantum-applications}\textbf{c}, or a single atom coupled to a whispering-gallery-mode microresonator~\cite{Sayrin2015PRX}. In both cases, the isolators were operated in a dissipative regime leading to decoherence, and they essentially behaved as classical devices. However, non-reciprocal devices based on chiral coupling have also been proposed in the quantum regime, such as optical circulators for single photons operated by a single quantum emitter~\cite{Sollner2015NNANO,Xia2014PRA}, cf. Fig.~\ref{Quantum-applications} \textbf{a}. By preparing the emitter in a superposition of spin ground states, this would enable preparing the circulator itself in a quantum superposition state where it both routes and does not route a light pulse at the same time. Another possibility is to exploit the intrinsic nonlinearity of the emitters at the single-photon level to achieve, e.g., photon number-dependent routing~\cite{Chang2014NPHOT, Javadi2015NCOM, Volz2012NPHOT}. These genuine quantum effects dramatically widen the range of potential applications and enable the development of photonic quantum circuits, where the chiral photon-emitter coupling is fully exploited for novel integrated quantum functionalities.

Recently, two quantum nonlinear devices have been demonstrated, which take advantage of the chiral coupling of a single rubidium atom to the modes of a whispering-gallery-mode microresonator~\cite{Shomroni2014Science, Rosenblum2016NPHOT}, cf. Fig.~\ref{Quantum-applications}\textbf{(b)}. In the first experiment, an optical switch controlled by a single photon was implemented~\cite{Shomroni2014Science} and, subsequently, the switch was used to extract a single photon from an optical pulse~\cite{Rosenblum2016NPHOT}. In both experiments, a single propagating photon transfers the atom from one ground state to another upon emission of a photon in the reverse direction. All subsequent photons are then simply transmitted. This scenario is analogous to the chiral absorber elaborated in Box~2, where the $\Lambda$-type level-structure of the rubidium atom ensures that the absorbed photon is re-emitted into the opposite direction of the waveguide. This mechanism is a key ingredient for a range of elementary quantum devices based on chiral quantum effects such as a nondestructive photon detector and a passive, heralded quantum memory. Also, the protocol can be employed to implement a $\sqrt{\rm SWAP}$ universal quantum gate for photons~\cite{Koshino2010PRA}.

Finally, chiral coupling in a waveguide opens new avenues for integrating quantum functionalities on a photonic chip. For instance, it may be used to map the spin state of a single electron or hole in a quantum dot to the propagation path of a photon enabling optical single-shot spin read out \cite{Young2015PRL}. Based on that, a layout of a deterministic photonic CNOT gate integrated on a chip was recently reported, for which the required experimental resources are within reach~\cite{Sollner2015NNANO}. Furthermore, a full architecture of distributed photonic quantum computing based on entanglement generation and local parity measurements has been proposed~\cite{Mahmoodian2016arxiv}.

\subsection{Chiral Quantum Many-Body Systems}

\begin{figure}
	\includegraphics[width=0.9\columnwidth]{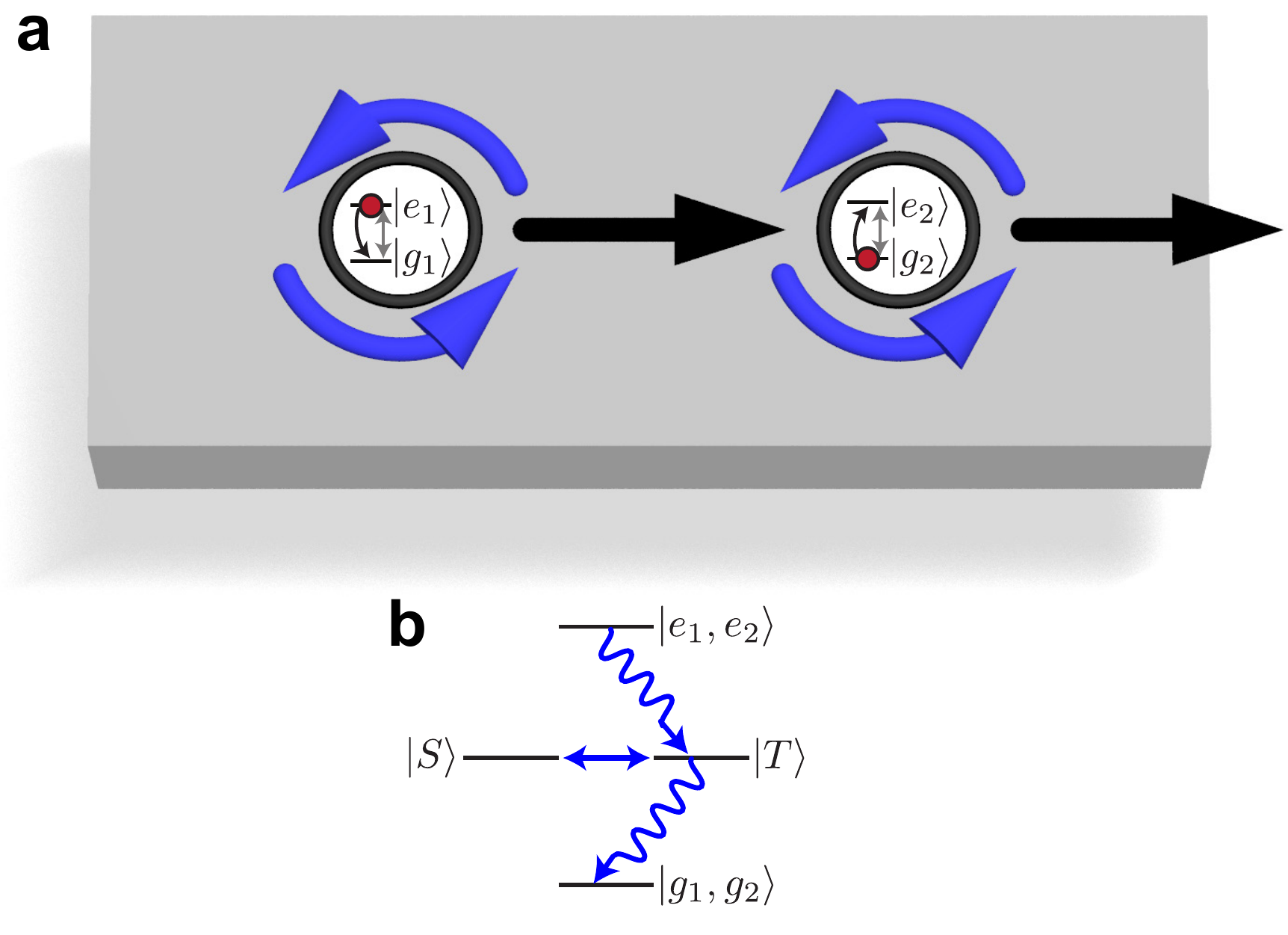}
	\caption{ Two 1D emitters mutually coupled via a chiral quantum channel in a photonic waveguide. The photons at the output port of the first emitter propagate to the input port of the second emitter, but not vice versa.  The corresponding emitter dynamics is illustrated in
(b). We show the collective states of the two two-level atoms (as two effective spin-$1/2$) in a basis of triplet $\ket{g_1.g_2}$, $\ket{T}\sim\ket{e_1,g_2}+\ket{g_1,e_2}$, $\ket{e_1,e_2}$ and singlet states $\ket{S}\sim\ket{e_1,g_2}-\ket{g_1,e_2}$, including their decay channels. For symmetric coupling, the singlet and triplet state correspond to subradiant and superradiant states, which are decoupled. In contrast, for chiral coupling $\ket{S}$ and $\ket{T}$ will be coupled. This underlies the formation of the driven pure quantum dimers as superposition of $\ket{g_1,g_2}$ and $\ket{S}$ (see text).
}
	\label{fig_Cascaded}
\end{figure}

1D quantum emitters coupled through chiral interaction to a waveguide provide a unique quantum many-body system. The emitters interact via exchange of photons \cite{Douglas:wy,Goban:2014eq,Hung:2016wy}, where the unidirectionality of photon emission into the waveguide leads to a chiral interaction between a pair of two-level systems
\cite{Stannigel2012NJP,Stannigel:2011ig,Ramos2014quantum,Pichler:2015jl}. To illustrate chiral interaction, consider a pair of emitters coupled to a 1D waveguide, c.f.~ Fig.~\ref{fig_Cascaded}. If the left emitter is initially in the excited state $\ket{e_1}$, while the right emitter is prepared in its ground state $\ket{g_2}$, the photon generated by decay of the first emitter can propagate to the right and be reabsorbed by the second emitter, resulting in a process $\ket{e_1}\ket{g_2} \rightarrow \ket{g_1}\ket{e_2}$.  For a symmetric emitter-waveguide coupling there is also the reverse process with a photon propagating from right to left. This is the familiar dipole-dipole coupling giving rise to a repeated exchange of excitations between the emitters, cf. Box~3.
For a chiral, or - in the extreme case - purely unidirectional coupling, emitters on the left can only excite emitters to the right, but not vice versa, breaking the left-right symmetry.
   The possibility of chiral interaction is intrinsically related to the connecting waveguide being infinitely extended, i.e., photons propagating along one direction are not reflected backwards. For a proper description it is thus essential to have an open quantum system, where  input ports facilitate injection of quantum signals, whereas photons leaving the output port never return.

From a formal quantum optical perspective, the chiral coupling of quantum systems leads to a cascaded quantum system \cite{Gardiner:1993cy,Carmichael:1993el}, and the term chiral quantum network  has been coined to describe a composite quantum system consisting of nodes connected by chiral quantum channels. 
A theory of quantum noise of cascaded quantum systems has been developed \cite{QWII} where the central idea is to treat the continuum of waveguide modes, representing the channel, as a quantum reservoir of bosonic excitations. The radiative coupling of emitters to the waveguide is modelled in the Markov approximation as a decay rate $\gamma$. In Box 3, the master equation for two emitters coupled by a chiral channel is presented. We emphasize that it is the combination of a coherent Hamiltonian and collective dissipative terms (describing superradiance), which provide a consistent description of chiral connections.

Chiral quantum networks have immediate applications for constructing quantum circuits for quantum--information processing \cite{Kimble:2008if}. With two-level emitters representing stationary qubits, and photons as flying qubits in a quantum network, the chiral light--matter coupling allows routing of photons in between the nodes \cite{Cirac:1997is,Stannigel:2011ig}. As a simple example, Fig.~\ref{fig_Cascaded} illustrates quantum state transfer of qubits by the protocol $(c_g \ket{g_1}+c_e \ket{e_1} ) \ket{g_2} \rightarrow \ket{g_1} (c_g \ket{g_2}+c_e \ket{e_2}) $, where an arbitrary superposition stored in emitter 1 is mapped to emitter 2.  Chiral couplings of emitters to the waveguide serve here as the mechanism to convert the first qubit with (ideally) unity efficiency to a right moving photonic qubit, and to guarantee complete reabsorption of this photon by the second emitter qubit. We note that this transfer protocol described in \cite{Cirac:1997is} requires {\em pulse shaping} of the flying qubit. This can be achieved by combining $\Lambda$-type three level systems with the two ground states representing the qubit and a Raman process involving a chiral coupling  \cite{Cirac:1997is}.  The advantage of having chiral coupling for the generation of emitter--emitter entanglement and the robustness towards imperfections has been recently studied  \cite{Cirac:1997is,GonzalezBallestero2015PRB}.

The directional photon exchange between emitters  with chiral interactions offers interesting novel perspective for many-body quantum dynamics.
As a basic example, consider the spontaneous emission of an ensemble of two-level emitters.
Due to the collective character of the bath, the emission differs strongly from that of independent emitters, an effect referred to as sub- and superradiance \cite{Dicke:1954bl}. For symmetric coupling, two emitters can share a single excitation in such a way that destructive interference of all emission amplitudes prevents the excited two-emitter system from decaying. In contrast, such sub-radiant behaviour is absent in the case of chiral coupling where unidirectional coupling implies that only one of the emitters ``knows'' about the presence of the other emitter, cf.~Fig.~\ref{fig_Cascaded} \textbf{a}.

By adding laser light to continuously excite the two-level emitter, a driven-dissipative quantum system is obtained, with a stream of scattered photons leaving through the output port. Such a driven system will eventually evolve to a dynamical equilibrium between pumping and light scattering, where the steady state of the emitters may be found from the cascaded master equation of Box 3. Quite remarkably, for chiral coupling, steady states exist in the form of  pure entangled many-atom states \cite{Stannigel:2012jk,Ramos:2014ut,Pichler:2015jl}, i.e.,~the initial density operator  $\rho$  describing the emitter evolves to a pure state $\rho \rightarrow \ket{\psi}\bra{\psi}$. For the example of two driven emitters in Fig.~\ref{fig_Cascaded} this pure state has the form of a quantum dimer $\ket{\psi}  \propto\ket{g_1}\ket{g_2} + \alpha \left( \ket{g_1}\ket{e_2} - \ket{e_1}\ket{g_2}  \right )$ involving EPR-type correlations between the pair of emitters. The physical origin behind this phenomenon is quantum interference: for chiral coupling, the stream of photons scattered from the first emitter can interfere destructively with the photons scattered from the second emitter, driving the system into a dark state of the cascaded quantum system, i.e.,~into a pure matter state with no light emerging through the output port. This formation of quantum dimers generalizes to many emitters, and can be interpreted as an example of a novel magnetic phase of pure quantum dimers, which emerges as a non-equilibrium phases in a many-body quantum system with chiral interaction. Another research direction investigates optical forces in a chiral channel and the process of self-organization of trapped atoms \cite{Eldredge2016arXiv}.

\begin{framed}
	\section{Box 3  | Cascaded Master Equations for Chiral Quantum Channels}
	If many independent 1D emitters are coupled to a waveguide the photons can mediate collective effects. Within the Born-Markov approximation, the photons can be eliminated and the dynamics of the reduced state of the emitters, $\rho$, is described by a master equation. For two emitters with a perfect unidirectional coupling to the waveguide ($\beta_+=1,\beta_-=0$), this leads to the simplest form of a so called {\em cascaded master equation} \cite{Gardiner:1993cy,Carmichael:1993el}, which reads
	\begin{align}\label{Cascaded_ME4}
		\dot\rho&=\mathcal{L}\rho\equiv-i(H_{\rm eff}\rho-\rho H_{\rm eff}^\dag)+\sigma\rho\sigma\dg.
	\end{align}
	and can be cast in explicit Lindblad form. The non-hermitian effective Hamiltonian associated with an evolution without quantum jumps has the form
	\begin{align}
		H_{\rm eff}=H_{\rm sys}-i\frac{\gamma}{2}\left(\sigma^+_1 \sigma^-_1+\sigma^+_2 \sigma^-_2+2\sigma^+_2 \sigma^-_1\right),\label{nonHermitian4}
	\end{align}
	It contains the bare Hamiltonians of the two individual emitters, $H_{\rm sys}=H_{1}+H_2$, and familiar non-hermitian terms leading to the decay of the excited states of the two emitters. The last term of \eqref{nonHermitian4} is special to the cascaded setting and breaks the symmetry between the two emitters: it is related to the process where the first emitters radiates a photon that is then absorbed by the second one. Finally, the last term of \eqref{Cascaded_ME4} is associated with quantum jumps corresponding to the detection of a photon at the output port of the waveguide. The corresponding jump operator $\sigma=\sigma^-_1+\sigma^-_2$ is a collective operator. The unidirectional flow of information in this cascaded setting is contained in the master equation \eqref{Cascaded_ME4}, as can be seen explicitly by calculating the equation of motions of the reduced states of the two individual emitters, $\rho_i\equiv \textrm{Tr}_i\{\rho\}$ ($i=1,2$). While the equation for $\rho_1$ is closed and independent of the state of the second atom, the reverse is not true.
	The cascaded master equation \eqref{Cascaded_ME4} can be generalized to include many emitters \cite{Stannigel:2012jk} and to the case of arbitrary $\beta_{+}$ and $\beta_{-}$ \cite{Ramos:2014ut,Pichler:2015jl}.
	
	It is worthwhile noting that in \eqref{Cascaded_ME4} the positions of the emitters, $x_i$, do not enter as any propagation phase acquired by the photons can always be gauged away in the cascaded setting. This is in contrast to the well-known {\em bidirectional case}, where the emitter dynamics explicitly depends on the phase that a photon with wavenumber $k$ picks up when travelling from $x_1$ to $x_2$:
	\begin{align}\label{bidirectional}
		\dot\rho&=-i[H_{\rm sys} +\gamma\sin(k |x_1-x_2|) (\sigma^+_1\sigma^-_2+\sigma^+_2\sigma^-_1),\rho]\nonumber\\&+2\gamma\sum_{i,j=1,2}\cos(k|x_i-x_j|)(\sigma_i^-\rho\sigma_j^+-\frac{1}{2}\{\sigma_i^+\sigma_j^-,\rho\}).
	\end{align}
	In this bidirectional setting, the 1D reservoir mediates (infinite-range) coherent dipole-dipole interactions and gives rise to super- and subradiance.
\end{framed}

\section{Outlook}

Modern photonic nanostructures are capable of enhancing light--matter coupling to such an extent that routinely single-quanta experiments can be carried out. Notably 1D waveguides offer the most elementary open quantum reservoir, which can be ingeniously engineered. Chiral light--matter interaction in 1D photonic systems is a natural consequence of the transverse confinement of electromagnetic fields, which leads to novel physical phenomena and applications. These include non-reciprocal quantum-photonic devices operating at the single--photon level and deterministic spin--photon interfaces potentially integrated on an optical chip. These are essential building blocks of future photonic quantum networks exploiting chiral quantum circuits. Extending chiral interaction to multiple emitters opens new avenues for quantum many--body physics.  New quantum phases of light and matter may be exploited by coupling photon qubits and emitter qubits by chiral interaction. In the regime of deterministic coupling, even a single photon suffices in saturating a single quantum emitter, which is the realm of single--photon nonlinear optics.  A chiral network of such efficiently coupled qubits interconnected by photons is likely to lead to photonic topological effects in fundamentally new and genuinely quantum regimes. Applications within quantum-information processing and for quantum simulation of complex systems are foreseen.

\bibliography{references}


\end{document}